\documentclass[%
reprint,
 amsmath,amssymb,
 aps,
 pra,
]{revtex4-1}

\usepackage{graphicx}
\usepackage{epsfig}
\usepackage{epstopdf}
\usepackage{subfigure,color}
\usepackage{dcolumn}
\usepackage{bm}
\usepackage{soul}
\begin{document}
	
\title{Instantaneous Radiation from Time-Varying Electric and Magnetic Dipoles}

\author{M.~S.~Mirmoosa$^{1}$, G.~A.~Ptitcyn$^2$, R.~Fleury$^1$, and S.~A.~Tretyakov$^2$}

\affiliation{$^1$Laboratory of Wave Engineering, {\'E}cole Polytechnique F{\'e}d{\'e}rale de Lausanne (EPFL), CH-1015 Lausanne, Switzerland\\$^2$Department of Electronics and Nanoengineering, School of Electrical Engineering, Aalto University, P.O.~Box 15500, FI-00076 Aalto, Finland} 

\begin{abstract}
Radiation from magnetic and electric dipole moments is a key subject in theory of electrodynamics. Although people treat the problem thoroughly in the context of frequency domain, the problem is still not well understood in the context of time domain, especially if dipole moments arbitrarily vary in time under action of external forces. Here, we scrutinize the {\it instantaneous} power radiated by magnetic and electric dipole moments, and report findings that are different from the conventional understanding of their instantaneous radiation found in textbooks. In contrast to the traditional far-field approach based on the Poynting vector, our analysis employs a near-field method based on the induced electromotive force, leading to corrective terms that are found to be consistent with time-domain numerical simulations, unlike previously reported expressions. Beyond its theoretical value, this work may also have significant impact in the field of time-varying metamaterials, especially in the study of radiation from subwavelength meta-atoms, scatterers and emitters that are temporally modulated.     
\end{abstract}

\maketitle 


\section{Introduction}
\label{sec:introduction}

Electromagnetic radiation is conventionally identified and studied by looking at the Poynting vector~\cite{Poynting} at large distances from the source. This perspective towards radiation is supported by the use of the Sommerfeld radiation condition~\cite{Sommerfeld,Schot} which is applied at those distances. As a consequence, one can say that the total energy per unit time radiated from a dipole moment can be obtained by using the following expression~\cite{Griffiths}:
\begin{equation}
P=\oint_S\Big[\mathbf{E}\times\mathbf{H}\Big]\cdot d\mathbf{S},
\end{equation}
in which $\mathbf{E}$ and $\mathbf{H}$ represent the ``far-field" components of the time-varying electric and magnetic fields, respectively, generated by the dipole moment. As is known, these ``far-field" components attenuate as $1/r$ in which $r$ is the distance from the origin where the dipole is located. In the literature, the electric current density inducing the dipole moment is usually assumed to be a time-harmonic function. Therefore, applying the method of phasors, the above equation is simplified and people are used to write the averaged energy per unit time over one cycle. We have
\begin{equation}
P_{\rm{av}}=\oint_S{1\over2}{\rm{Re}}\Big[\overline{\mathbf{E}}\times\overline{\mathbf{H}}^\ast\Big]\cdot d\mathbf{S},
\label{eq:2}
\end{equation} 
in which $^\ast$ denotes the complex conjugate and ${\rm{Re[\, ]}}$ gives the real part. Here, $\overline{\mathbf{E}}$ and $\overline{\mathbf{H}}$ are the ``far-field" components of the complex phasors of the electric and magnetic fields in the frequency domain. 

In this paper we introduce an alternative approach to study radiation from dipoles, that uses the fields at the vicinity of the dipole (notice that the near-field components of the fields have a singularity at $r=0$). This point of view, which we develop not only in frequency domain, but also directly in time domain, has not been explicitly explored in the literature, and leads to interesting new results and insights. Counter-intuitively, we demonstrate that the non-singular component for the fields at the location of the dipole determines not only the time-averaged power given by \eqref{eq:2}, but also additional instantaneous exchange of power between the dipole and the fields. This non-singular component is proportional to the second and the third derivatives of the electric current regarding electric and magnetic dipole moments, respectively. Interestingly, this non-singular component is contributing to the induced electromotive force which results in radiation. We compare the time-varying induced electromotive force with the Lorentz-friction force exerted on one single accelerated electron and explicitly show the fundamental resemblance between these two classical concepts. In the end, we show some simulated results confirming our theoretical expectations about the instantaneous radiation. This work may have impact on the study of transient description of antennas~\cite{Papas,Miller,Maloney,Smith,Smith_teaching,Smith_Hertel}, transient description of the reactive power around the dipole~\cite{Schantz1,Kaiser,Valagiannopoulos_Alu,Vandenbosch1,Vandenbosch2}, time-modulated scatterers~\cite{ptitcyn}, and also time-modulated metamaterials (see e.g.~\cite{Caloz1111,Caloz2222,Fleurypra}).

The paper is organized as follows: In Sections~\ref{sec:mag} and \ref{sec:elec}, we show the corresponding analytical derivations regarding magnetic and electric dipole moments, respectively. In Section~\ref{sec:numerical}, we demonstrate the full-wave simulation results confirming the analytical results achieved in the previous Sections. In Section~\ref{sec:nstdim}, we repeat our derivations assuming that the electric and magnetic dipole moments have arbitrary temporal variation (not only time harmonic), and finally in Section~\ref{sec:conclusion}, we conclude the paper. 

\section{Magnetic dipole}
\label{sec:mag}
Let us consider a radiating loop that has a small radius compared to the wavelength. The loop is centred at the origin on the $xy$-plane. In the frequency domain, the theta component of the generated complex magnetic field is given by~\cite{Balanis}
\begin{equation}
\overline{H}_\theta=-{(ka)^2I_0\over4r}\Big[1+{1\over jkr}-{1\over(kr)^2}\Big]\sin\theta\exp[-jkr],
\label{eq:expexpmag}
\end{equation}
in which $k$ is the free-space wave number, $a$ is the loop radius, and $I_0$ represents the electric current carried by the loop. We are interested in contemplating the magnetic field extremely close to the loop in order to find electromotive force induced by the fields into the antenna. For this, we employ the Taylor series for the exponential function:
\begin{equation}
\exp[-jkr]=1-jkr-{k^2r^2\over2}+j{k^3r^3\over6}+\dots
\label{eq:expansion}
\end{equation}  
Later we will see that for our purpose it is enough to keep the first four members of the series. We substitute the above expression into  Eq.~\eqref{eq:expexpmag} and rewrite the magnetic field. Multiplying by the free-space permeability, the magnetic flux density reduces to 
\begin{equation}
\overline{B}_\theta=jk^3{\mu_0a^2I_0\over6}\Big[1+j{3\over4kr}-j{3\over2(kr)^3}\Big]\sin\theta.
\label{eqbafexp}
\end{equation} 
We must consider the flux density normal to the $xy$-plane (where the loop is located), meaning that the elevation angle $\theta$ is equal to $\pi/2$. Subsequently, the total flux is readily found by multiplication of the flux density by the surface of the loop. Therefore, we have
\begin{equation}
\overline{\phi}=\overline{\mathbf{B}}\vert_{\theta={\pi\over2}}\cdot S\mathbf{a}_z=jk^3{\mu_0(\pi a^2)^2I_0\over6\pi}\Big[1+j{3\over4kr}-j{3\over2(kr)^3}\Big].
\end{equation}  
Here, $\mathbf{a}_z$ is the unit vector normal to the loop plane and $S$ is the loop area. 
The real part of this complex amplitude is associated with the reactive power which describes the stored energy near the radiator, and it decreases as the distance grows. In contrast, the imaginary part corresponds to the active power radiated into space, and it is not inversely proportional to the distance. Consequently, we focus the attention on the imaginary part: 
\begin{equation}
\overline{\phi}_{\rm{imag}}=jk^3{\mu_0(\pi a^2)^2I_0\over6\pi}.
\end{equation}  
We assume that $I_0$ is a real value. Intriguingly, there is no singularity for this component of the flux, while the other component (the real part) is singular. This indicates that the radiated power is finite at any moment of time although the source can have an arbitrarily small size and the fields are singular. After finding the non-singular component, we do not continue with the frequency analysis, and we move to the time domain. Pondering about the above equation, we discern that the flux is associated with the third derivative of the electric current. Why? Because we have the wave number in power three. Remind that for an arbitrary function $f(t)$ having the Fourier transform $F(\omega)$, we have  
\begin{equation}
{d^3f(t)\over dt^3}\longrightarrow -j\omega^3F(\omega),
\end{equation}      
in which $\omega=kc$ ($c$ is the speed of light) is the angular frequency. Here, the arrow means the Fourier transform. Thus, according to this explanation, the instantaneous flux associated with the radiated power is given by 
\begin{equation}
\phi_{\rm{imag}}(t)=-{\mu_0(\pi a^2)^2\over6\pi c^3}{d^3i(t)\over dt^3}.
\end{equation}
Faraday's law will help us to derive the electromotive force induced due to the temporal variation of the flux. Keeping this in mind, finally the instantaneous radiated power can be obtained through 
\begin{equation}
P_{\rm{m}}(t)=v(t)\cdot i(t)=-{d\phi_{\rm{imag}}(t)\over dt}\cdot i(t).
\end{equation}     
Since the radiated power is proportional to the first derivative of the flux, the radiated power is consequently proportional to the fourth derivative of the electric current and also to the electric current itself. In terms of the magnetic moment, the same conclusion is true since the magnetic moment is related to the electric current as $m(t)=(\pi a^2)i(t)$. After doing some algebraic manipulations, we can write 
\begin{equation}
P_{\rm{m}}(t)={\mu_0\over6\pi c^3}m(t)\cdot{d^4m(t)\over dt^4}.
\label{eq:pomdm}
\end{equation} 

This is a key result, and we now discuss it. In the context of time-harmonic fields, if the magnetic moment is described by $m(t)=m_0\cos(\omega t)$, the radiated power is found as $P_{\rm{m}}(t)=(\mu_0/6\pi c^3)\omega^4m_0^2\cos^2\omega t$. If we write the instantaneous power in terms of the electric current and calculate the time-averaged radiated power (over one cycle), we see that
\begin{equation}
P^{\rm{av}}_{\rm{m}}={\mu_0\over12\pi c^3}\omega^4(\pi a^2)I_0^2.
\end{equation}
Based on the terminology of antenna engineering, the averaged radiated power is expressed through the concept of the radiation resistance. Reminding that $\mu_0c=120\pi$, this resistance is found as 
\begin{equation}
R_{\rm{m}}={2P^{\rm{av}}_{\rm{m}}\over I_0^2}=20\pi^2\Big({C\over\lambda}\Big)^4,
\end{equation}   
in which $C=2\pi a$ is the circumference of the loop. This expression is very well known for antenna engineers (see e.g.~Ref.~\cite{Balanis}). 

Equation~\eqref{eq:pomdm} is more complex. We know that
\begin{equation}
m\cdot{d^4m\over dt^4}={d\over dt}\Big[m\cdot{d^3m\over dt^3}\Big]-{dm\over dt}\cdot{d^3m\over dt^3},
\label{eq:4der}
\end{equation}
where 
\begin{equation}
{dm\over dt}\cdot{d^3m\over dt^3}={d\over dt}\Big[{dm\over dt}\cdot{d^2m\over dt^2}\Big]-\left({d^2m\over dt^2}\right)^2.
\end{equation}
As a consequence, Eq.~\eqref{eq:4der} reduces to
\begin{equation}
m\cdot{d^4m\over dt^4}=\left({d^2m\over dt^2}\right)^2+{d\over dt}\Big[m\cdot{d^3m\over dt^3}-{dm\over dt}\cdot{d^2m\over dt^2}\Big].
\end{equation}
As far as we know, the last term (in the square brackets), is absent in all textbooks. However, only in the context of time-harmonic fields it averages to zero. In the time domain, this term is actually crucial for correct satisfaction of the instantaneous power balance, as we demonstrate in Section~IV. This critical point is even more important for the electric dipole, for which neglecting the second term gives dramatically different and wrong results.

Let us go deeper and investigate the magnetic-dipole radiation from one single electron rotating with acceleration. Naoki Itoh, in Ref.~\cite{Itoh}, wrote the corresponding electric field generated by such electron (via the expansion method, the fourth-order term) as 
\begin{equation}
\mathbf{E}={\mu_0\over12\pi c^3}\mathbf{R}\times{d^4\mathbf{m}\over dt^4},
\end{equation}  
where $\mathbf{R}$ is the position vector of the observation point. The magnetic moment is expressed in terms of the velocity as 
\begin{equation}
\mathbf{m}={e\over2}\mathbf{R}\times\mathbf{v}.
\end{equation}
The force due to the electric field acting on the electron is $\mathbf{F}=e\mathbf{E}$. Therefore, the force finally reduces to \begin{equation}
\mathbf{F}=-e\Bigg[{\mu_0\over6\pi c^3}\left({R\over2}\right)^2{d^4e\mathbf{v}\over dt^4}\Bigg].
\end{equation}
This expression is indeed quite similar to the electromotive force that was found for the magnetic dipole: 
\begin{equation}
v(t)=-{d\phi_{\rm{imag}}(t)\over dt}={\mu_0\over6\pi c^3}(\pi a^2)^2{d^4i(t)\over dt^4}.
\label{eq:emfmd}
\end{equation} 


\section{Electric dipole} 
\label{sec:elec}
Let us consider a radiating Hertzian dipole which has a dipole moment (in the frequency domain) along the $z$ axis: $\overline{\mathbf{p}}=(I_0l/j\omega)\mathbf{a}_z$ in which $l$ is the length of the dipole. The electric field corresponding to the elevation angle equal to $\pi/2$ is also parallel to the $z$ axis and can be expressed as~(e.g., \cite{Balanis}) 
\begin{equation}
\overline{\mathbf{E}}=-j\eta{kI_0l\over4\pi r}\Big[1+{1\over jkr}-{1\over(kr)^2}\Big]\exp(-jkr)\mathbf{a}_z,
\end{equation} 
where $\eta$ is the free-space intrinsic impedance. Similar to what we did for the magnetic dipole, we look at the field at the vicinity of the Hertzian dipole and expand the exponential function up to the 4th term, see Eq.~\eqref{eq:expansion}. The electric field is simplified and the nonsingular component of the field emerges (${\mathbf{E}}_{\rm{ns}}$). This component is given by
\begin{equation}
\overline{\mathbf{E}}_{\rm{ns}}=(jk)^2{\eta I_0l\over6\pi}\mathbf{a}_z.
\end{equation}
What does the second power of the wave number mean? It shows that the electric field is proportional to the second derivative of the electric current in time. Note that the wave number is the angular frequency divided by the speed of light. Ergo, the inverse Fourier transform gives 
\begin{equation} 
\mathbf{E}_{\rm{ns}}(t)={\mu_0\over6\pi c}l{d^2i(t)\over dt^2}\mathbf{a}_z.
\label{eq:nselhar}
\end{equation}
Here, we have used the equality $\eta=\mu_0c$. Now, since we have found the electric field, we can readily derive the induced electromotive force through 
\begin{equation}
v_{\rm{ED}}(t)=-\int\mathbf{E}\cdot d\mathbf{l}=-{\mu_0\over6\pi c}l^2{d^2i(t)\over dt^2}.
\label{eq:emfed}
\end{equation}	
There are two explicit differences between the electromotive force for the electric dipole and for the magnetic dipole. Firstly, $v(t)$ was inversely proportional to the third power of the speed of light while $v_{\rm{ED}}(t)$ is inversely proportional to the first power of $c$. Furthermore, $v(t)$ is related to the fourth derivative of the electric current but $v_{\rm{ED}}(t)$ is related to the second derivative of the current. Comparing Eqs.~\eqref{eq:emfmd} and \eqref{eq:emfed} is intriguing and informative. Area is substituted by length, $1/c^3$ by $1/c$, and the 4th derivative by the 2nd derivative. Knowing the electromotive force, we  readily derive the instantaneous radiated power using
\begin{equation}
P_{\rm{e}}(t)=v_{\rm{ED}}(t)\cdot i(t).
\label{eq:vipower}
\end{equation}     

Our purpose is to write the radiated power based on the electric dipole moment. We use the fact that the electric current multiplied by the length of the dipole is the time derivative of the electric dipole moment: $i(t)l=dp(t)/dt$. Accordingly, we eventually express the radiated power as 
\begin{equation}
P_{\rm{e}}(t)=-{\mu_0\over6\pi c}{dp(t)\over dt}\cdot{d^3p(t)\over dt^3}.
\label{eq:ededed}
\end{equation}
For a time-harmonic moment, $p(t)=p_0\cos(\omega t)$, the radiated power is identical with $P_{\rm{e}}(t)=[{\mu_0/6\pi c}]p_0^2\omega^4\sin^2\omega t$. Consequently, the averaged radiated power is half of the amplitude, and we can write the radiation resistance of the radiator (in the frequency domain) as \begin{equation}
R_{\rm{e}}={2P^{\rm{av}}_{\rm{e}}\over\vert I_0\vert^2}={\mu_0\over6\pi c}(\omega l)^2=80\pi^2\left({l\over\lambda}\right)^2,
\end{equation} 
which is the known expression in all antenna books, e.g.~Ref.~\cite{Balanis} (reminding that $\eta=120\pi$). 

The radiation reaction force acting on the accelerated electron which has a velocity $\mathbf{v}$ is~\cite{Landau}
\begin{equation}
\mathbf{F}_{\rm{e}}=-e\left(-{\mu_0\over6\pi c}{d^2e\mathbf{v}\over dt^2}\right).
\end{equation}
Again, similarly to the radiating magnetic moment, the electromotive force found above shows a strong link to the force exerted on one electron. The radiated power from an accelerated electron is given by $P_{\rm{electron}}=-\mathbf{F}_{\rm{e}}\cdot\mathbf{v}$, which is similar to  Eq.~\eqref{eq:vipower} used for the Hertzian dipole. The total power radiated by the dipole is the product of the electromotive force and the electric current (that is related to the charge velocity). 


\begin{figure*}[!t]
\centerline{\includegraphics[width=15cm]{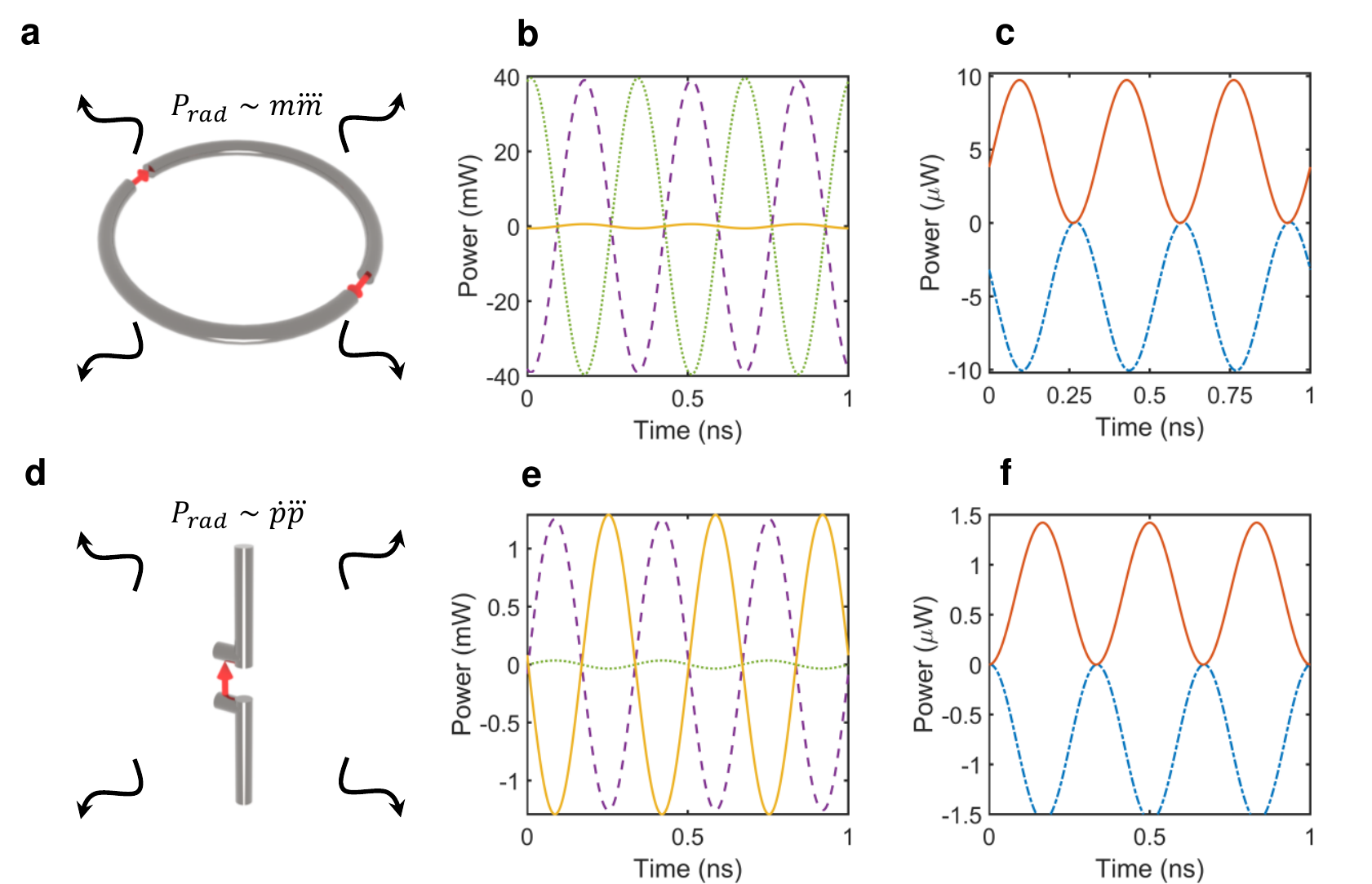}}
\caption{(a)--(c) The instantaneous powers for the radiating magnetic dipole. (d)--(f) The  instantaneous powers for the radiating electric dipole. The orange solid curve and the green dotted curve represent the electric and magnetic energy per unit time, respectively. Also, the purple dashed curve indicates the supplied energy per unit time. Finally, the red solid curve corresponds to the instantaneous radiated power and the blue dot-dashed curve determines the sum of the electric, magnetic, and supplied energy per unit time. The radius of the loop is 3 mm and the length of the cylindrical wire is 15 mm. Both the loop and the wire are made of perfect conductor, and are much smaller than the operating wavelength that is 200 mm. The wire is fed in the center and the loop is fed by two differential sources such that the electric dipole mode vanishes. The sources are determined by the red arrows.}
\label{figureone}
\end{figure*}

\section{Numerical results} 
\label{sec:numerical}
To validate the theoretical results, we did full-wave simulations employing COMSOL Multiphysics software. As shown in Fig.~\ref{figureone}(a), we excite an electrically small loop, made of perfect conductor, with two simultaneous sources clockwise. The reason for that is to cancel the extra higher modes mainly including the electric dipole moment. Therefore, the magnetic dipole moment is only excited. We employ the instantaneous power balance: 
\begin{equation}
\Sigma P(t)=0,
\label{eq:pbins}
\end{equation}
in which $\Sigma P(t)$ includes the supplied power, the reactive power (the electric and magnetic energy per unit time), and the radiated power. Figure~\ref{figureone}(b) shows the reactive and supplied power. Based on the above equation, we must have 
\begin{equation}
P_{\rm{supplied}}(t)+P_{\rm{reactive}}(t)=-P_{\rm{radiative}}(t).
\end{equation}
The left side is extracted from the simulator and compared with the right side calculated theoretically according to the Eq.~\eqref{eq:pomdm}. Figure~\ref{figureone}(c) demonstrates that the simulated and theoretical results are in agreement. We have also simulated different loops with different radii. Until the radius is electrically small enough, we have found that the error between simulations and theory is negligible. However, increasing the radius causes that higher-order modes become significant and they must be taken into account in the power balance, which is not straightforward. 

For the electric dipole moment, we have simulated an electrically small wire, shown in Fig.~\ref{figureone}(d), that is excited by one source located at the center. The supplied and the reactive powers are shown in Fig.~\ref{figureone}(e), and the comparison between the summation of those powers with the radiated power is indicated by Fig.~\ref{figureone}(f). As it is seen, if we use the theoretical expression given by Eq.~\eqref{eq:ededed}, the theory and the simulated results are in agreement. However, similar to the magnetic moment, one can say that Eq.~\eqref{eq:ededed} can be modified based on the following consideration:
\begin{equation}
{dp\over dt}\cdot{d^3p\over dt^3}={d\over dt}\Big[{dp\over dt}\cdot{d^2p\over dt^2}\Big]-\left({d^2p\over dt^2}\right)^2,
\end{equation}
interpreting that the first term in the right-hand side is not relevant to the radiated power because it is related to the stored electromagnetic energy. In other words, one can say that the correct form of Eq.~\eqref{eq:ededed} is $P_{\rm{e}}(t)=(\mu_0/6\pi c) (d^2p/dt^2)^2$. This conceptual issue is, indeed, challenging. But regardless of the conundrum that we consider the additional term in the radiated or reactive power, the fact is that we cannot certainly neglect the additional term $d/dt[({dp/dt})\cdot({d^2p/dt^2})]$ calculating the  instantaneous power balance. The simulated results completely confirm that neglecting this term causes violation of Eq.~\eqref{eq:pbins}. Put another way, $\Sigma P(t)$ is only zero if we take into account the additional term. It does not matter which interpretation we choose, this term is definitely there in the power balance.  


\section{Time-varying electric and magnetic dipoles}
\label{sec:nstdim}

Although the derivations in previous sections are valid for instantaneous radiation, we basically focused on the fields which resulted from time-harmonic oscillations. However, we can arrive to general conclusions by carefully looking at the general expressions given for the time-varying electric and magnetic fields of arbitrary nonstatic electric and magnetic dipoles. In Refs.~\cite{Griffiths,Schantz,Farina}, the time-varying electric field was derived and written for the electric dipole moment taking into account the retardation effect. According to those references, we have      
\begin{equation}
\begin{split}
\mathbf{E(r},t)=&-{\mu_0\over4\pi}\Bigg[{\ddot{\mathbf{p}}-\mathbf{a}_r(\mathbf{a}_r\cdot\ddot{\mathbf{p}})\over r}\cr
&+c^2{\Big[\mathbf{p}+(r/c)\dot{\mathbf{p}}\Big]-3\mathbf{a}_r\Big[\mathbf{a}_r\cdot[\mathbf{p}+(r/c)\dot{\mathbf{p}}]\Big]\over r^3}\Bigg]\cr
&-{1\over3\epsilon_0}\mathbf{p}\delta^3(\mathbf{r}),
\end{split}
\label{eq:elextv}
\end{equation}
where $\delta^3(\mathbf{r})$ denotes the three-dimensional Dirac delta function. In Eq.~\eqref{eq:elextv}, the electric field at time $t$ is due to the dipole moment at $t_{\rm{r}}=t-r/c$ (retardation effect). In other words, 
\begin{equation}
\mathbf{p}(t_{\rm{r}})=\mathbf{p}(t-{r\over c}).
\end{equation}
Applying the expansion technique, the dipole moment in the above is readily simplified to 
\begin{equation}
\mathbf{p}(t-{r\over c})=\mathbf{p}(t)-{r\over c}\dot{\mathbf{p}}(t)+{1\over2}({r\over c})^2\ddot{\mathbf{p}}(t)-{1\over6}({r\over c})^3\dddot{\mathbf{p}}(t)+...
\label{eq:expdip}
\end{equation}
Let us consider the limit as $r$ approaches zero ($r\rightarrow0$). Taking this limit in Eq.~\eqref{eq:elextv}, we can neglect the $\mathbf{a}_r$-component of the electric field and only consider the component which is parallel to the electric dipole moment: 
\begin{equation}
\mathbf{E}_{\rm{p}}(\mathbf{r},t)=-{\mu_0\over4\pi}\Bigg[{\ddot{\mathbf{p}}\over r}+c^2{\Big[\mathbf{p}+(r/c)\dot{\mathbf{p}}\Big]\over r^3}\Bigg]-{1\over3\epsilon_0}\mathbf{p}\delta^3(\mathbf{r}),
\end{equation}
and if we replace Eq.~\eqref{eq:expdip}, after some algebraic manipulations we deduce the non-singular component of the electric field at the dipole location as: 
\begin{equation}
\mathbf{E}_{\rm{ns}}(\mathbf{r},t)={\mu_0\over6\pi c}{d^3\mathbf{p}(t)\over dt^3}.
\end{equation}
This expression is exactly identical with the expression given by Eq.~\eqref{eq:nselhar} (reminding that the electric current multiplied by the electric dipole length is equal to the first time derivative of the electric dipole moment). Thus, the temporal electromotive force and the radiated power are subsequently the same as what we derived before. 

Regarding magnetic dipole moments, the magnetic flux density is expressed as~\cite{Griffiths,Schantz,Farina}
\begin{equation}
\begin{split}
\mathbf{B(r},t)=&-{\mu_0\over4\pi}\Bigg[{\ddot{\mathbf{m}}-\mathbf{a}_r(\mathbf{a}_r\cdot\ddot{\mathbf{m}})\over c^2r}\cr
&+{\Big[\mathbf{m}+(r/c)\dot{\mathbf{m}}\Big]-3\mathbf{a}_r\Big[\mathbf{a}_r\cdot[\mathbf{m}+(r/c)\dot{\mathbf{m}}]\Big]\over r^3}\Bigg]\cr
&+{2\mu_0\over3}\mathbf{m}\delta^3(\mathbf{r}),
\end{split}
\end{equation}
that is in duality with the electric dipole moment, as expected. Repeating the same procedure that we did for electric dipoles, we obtain the following non-singular component of the magnetic flux density that is parallel to the magnetic moment: 
\begin{equation}
\mathbf{B}_{\rm{ns}}(\mathbf{r},t)={\mu_0\over6\pi c^3}{d^3\mathbf{m}(t)\over dt^3}.
\end{equation}
Comparing the above equation and Eq.~\eqref{eqbafexp}, we see that the corresponding non-singular components are quite similar (in Eq.~\eqref{eqbafexp}, we wrote the magnetic flux density based on the electric current and the expression is before the inverse Fourier transform). 

Consequently, from the above two equations giving the non-singular components of the electric and magnetic fields corresponding to the electric and magnetic moments, respectively, we understand that they are proportional to the third derivative of the dipole moment. However, since for the magnetic dipole the electromotive force is associated with the time derivative of the magnetic flux, the radiated power is proportional to the fourth derivative of the dipole moment in contrary to the electric dipole whose radiated power is proportional to third time derivative of the dipole moment.   


\section{Conclusions}
Using the concept of electromotive force and contemplating the electromagnetic fields at the vicinity of the dipole moments, we obtained the instantaneous radiated power. We observed that the power is proportional to the magnetic moment and its fourth time derivative. However, regarding electric moment, the power is proportional to the first and the third derivatives of the moment. These theoretical expressions were confirmed by several simulations of a loop and a cylindrical wire excited by time-harmonic sources. Also, we compared the electromotive force with the Lorentz-friction force and showed the similarity of these two concepts. Furthermore, based on the expansion technique, we analytically showed that our theoretical conclusions are not only valid for time-harmonic oscillation but they are also valid for any nonstatic moments having arbitrary temporal variation. This work may have strong potential to influence engineers interested in antenna engineering and scattering from subwavelength particles both in microwave and optical regimes. A possible future direction would be to consider an arbitrary volume current distribution and derive the instantaneous radiation including higher modes as well.        
\label{sec:conclusion}


\end{document}